\documentclass[10pt,letterpaper,aps,onecolumn,superscriptaddress,floatfix,notitlepage]{revtex4-1}
\usepackage[latin1]{inputenc}
\usepackage{natbib,color}
\usepackage{amsmath}
\usepackage{amsfonts}
\usepackage{amssymb}
\usepackage{bm}
\usepackage{graphicx}
\usepackage{graphicx}
\begin{document}
	\title{Andreev reflections and the quantum physics of black holes} \author{Sreenath K. Manikandan}
	\email{skizhakk@ur.rochester.edu}
	\affiliation{Department of Physics and Astronomy, University of Rochester, Rochester, NY 14627, USA}
	\affiliation{Center for Coherence and Quantum Optics, University of Rochester, Rochester, NY 14627, USA}   
	\author{Andrew N. Jordan}
	\email{jordan@pas.rochester.edu}
	\affiliation{Department of Physics and Astronomy, University of Rochester, Rochester, NY 14627, USA}
	\affiliation{Center for Coherence and Quantum Optics, University of Rochester, Rochester, NY 14627, USA}    
	\affiliation{Center for Quantum Studies, Chapman University, Orange, CA, USA, 92866}
	\date{\today}
	
	\begin{abstract}
		We establish an analogy between superconductor-metal interfaces and the quantum physics of a black hole, using the proximity effect. We show that the metal-superconductor interface can be thought of as an event horizon and Andreev reflection from the interface is analogous to the Hawking radiation in black holes. We describe quantum information transfer in Andreev reflection with a final state projection model similar to the Horowitz-Maldacena model for black hole evaporation.  We also propose the Andreev reflection-analogue of Hayden and Preskill's description of a black hole final state, where the black hole is described as an information mirror. The analogy between Crossed Andreev Reflections and Einstein-Rosen bridges is discussed: our proposal gives a precise mechanism for the apparent loss of quantum information in a black hole by the process of nonlocal Andreev reflection, transferring the quantum information through a wormhole and into another universe. Given these established connections, we conjecture that the final quantum state of a black hole is exactly the same as the ground state wavefunction of the superconductor/superfluid in the Bardeen-Cooper-Schrieffer (BCS) theory of superconductivity; in particular, the infalling matter and the infalling Hawking quanta, described in the Horowitz-Maldacena model, forms a Cooper pair-like singlet state inside the black hole. A black hole evaporating and shrinking in size can be thought of as the analogue of Andreev reflection by a hole where the superconductor loses a Cooper pair. Our model does not suffer from the black hole information problem since Andreev reflection is unitary. We also relate the thermodynamic properties of a black hole to that of a superconductor, and propose an experiment which can demonstrate the negative specific heat feature of black holes in a growing/evaporating condensate.
	\end{abstract}
	
	\maketitle
	\section{Introduction}
	\label{S:1}
	In this paper we study an analogy between the superconducting phase in superconductors and a black hole which is approaching its final state~\cite{horowitz,preskill}. A black hole reaches the ``half way point'' when it has already radiated half the initial entropy and cannot accept information anymore~\cite{preskill}. The information is reflected via Hawking radiation~\cite{Hawking1975} very quickly, and it is speculated that quantum theories of gravity are necessary to understand the process~\cite{preskill,horowitz,projection,infopara}. Attempts to understand the quantum physics of black holes have led to interesting black hole analogies proposed and observed in a variety of experiments, including lasers~\cite{bec,laser}, rapid change of dielectric constant in waveguides~\cite{waveguide}, and time-varying refractive index of a medium~\cite{Refraction}, moving plasma mirrors~\cite{plasma}, sonic systems~\cite{sound1,sound2} and  Bose-Einstein condensates~\cite{zapata2011resonant}. Hawking radiation analogues in these experimental systems can be understood as a process where the incoming modes are converted to the outgoing modes, as discussed by Jacobson~\cite{bhmodes}. The connection to gravity could be made, for example, by considering fluids in motion that creates a ``sonic horizon"~\cite{zapata2011resonant}, and it has been shown by Unruh in 1981 that the behavior of sound waves in a hypersonic fluid background is the analogue of the behavior of scalar waves in a black hole spacetime~\cite{sound1,sound2}. These considerations have been extended to gravity analogues in both fermionic and bosonic superfluids~\cite{fischer2001thermal}. In this paper, we present a different approach to this problem that can be thought as a solid state analogue of a black hole; we build up on the original considerations of mode conversion in condensed matter systems introduced by Andreev for normal metal-superconductor interfaces~\cite{andreev,Pannet,spintron,artemenko1978excess,artemenko1979theory,artemenko1979excess,zaitsev1980theory}, and relate that to Hawking radiation based on the final state projection models of black hole evaporation~\cite{horowitz,projection}. Further, we use this analogy to propose a solid state quantum analogue of several phenomena in quantum gravitational physics, such as the 
	information reflection from black holes and the wormhole travel of information. We note that the analogy between Andreev reflections and Hawking radiation is a possibility that was first pointed out by Jacobson~\cite{bhmodes}, even before the development of the final state projection approach~\cite{projection,horowitz} we take in this manuscript. Hence our considerations go beyond and relate the quantum physics of Andreev reflections to unitary models of black hole evaporation from an information theory point of view. We particularly focus on two models: The Horowitz-Maldacena model for black hole evaporation~\cite{horowitz} and Hayden and Preskill's black hole information mirror model~\cite{preskill}, and show that the mode conversion process in black hole evaporation~\cite{bhmodes,horowitz,preskill} is analogous to the mode conversion in Andreev reflections, and describe the latter as a deterministic quantum teleportation~\cite{CTC}. The mode conversion process in bosonic condensates can also be understood via a bosonic version of Andreev reflection, as studied in~\cite{zapata2011resonant,zapata2009andreev}.
	\begin{figure}
		\includegraphics[scale=0.5]{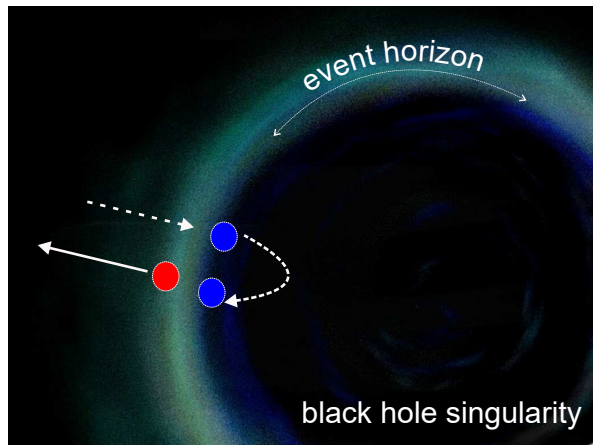}
		\caption{Hawking Radiation from a black hole in the Horowitz-Maldacena model for black hole evaporation: Virtual particle-antiparticle pairs exist at the event horizon of a black hole.  An incoming particle pairs with one of the virtual particle, and enters the black hole interior.  This process results in the ejection of a quasiparticle which can escape to infinity.\label{cosmos}}
	\end{figure}
	
	Hawking radiation, in the Horowitz-Maldacena model for black hole evaporation, is mediated by particle-antiparticle pairs created at the event horizon in close proximity to the black hole. One among the two particles from the pair falls into the black hole with the infalling matter and the other particle carries away the quantum information encoded in the infalling matter to infinity. The information transfer in black hole evaporation from the interface is also consistent with the holography principle~\cite{hologram}, which assumes that the information falling into the black holes can be extracted from the surface.  
	
	The theory of superconductivity and the theory of black holes contain many similarities, and this has lead to interesting new perspectives of investigating one in light of the parallels with the other. The theory of holographic superconductors is an example where one benefits from the gauge/gravity duality to provide a dual gravitational description for the superconductors~\cite{holo2,gaugegravity,holosc}. There is also a sense in which a class of black holes are called \textit{superconducting}, when the black holes undergo a phase transition near the event horizon with a broken symmetry that is $U(1)$, similar to the spontaneously broken $U(1)$ gauge symmetry for superconductors in the Landau-Ginzburg theory~\cite{london,scbh,scbh2}. 
	
	We first review the statement of the problem of unitarity of black hole evaporation (Susskind et al.,~\cite{complementarity}, also see ~\cite{infopara}), where to begin with, one assumes that a local description of quantum field theory exists and the universe can be factorized in to subsystems as desired for the conventional descriptions of quantum information dynamics and entanglement~\cite{horowitz,infopara,projection}. The notion of different instances of time is introduced by considering constant time slices of the spacetime manifold, called Cauchy surfaces. Let a constant time slice of the spacetime manifold prior to the formation of the black hole be the Cauchy surface $\Gamma$. Formation of an event horizon can be described similarly by another Cauchy surface $\Gamma'$ which is a product of two causally disconnected Cauchy surfaces, $\Gamma_{in}$ and $\Gamma_{ex}$,
	\begin{equation}
	\Gamma' = \Gamma_{in}\times\Gamma_{ex},\end{equation}
	for the interior and exterior of the black hole, respectively. The assumption of validity of local quantum field theory, linearity and unitarity of quantum mechanics together imply that the time evolution between Cauchy surfaces is locally described by a linear Schr\"{o}dinger equation. Under these assumptions, for an observer who is exterior to the black hole, the Hilbert space of the interior and exterior of the black hole is a product since the two surfaces ($\Gamma_{in},~\Gamma_{ex}$) are causally disconnected,
	\begin{equation}
	H=H_{in}\otimes H_{ex}.
	\end{equation}
	The Cauchy surface after the evaporation of the black hole is labeled by $\Gamma''$. We proceed with the assumption that this evolution  $\Gamma\rightarrow\Gamma'\rightarrow\Gamma''$ is linear and unitary. This implies that a pure wavefunction $\Psi(\Gamma)$ defined on the time slice $\Gamma$ is related to $\Psi(\Gamma'')$ defined on $\Gamma''$  via a unitary transformation $\mathcal{U}$,
	\begin{equation}
	\Psi(\Gamma'')=\mathcal{U}~\Psi(\Gamma).
	\end{equation}
	The Cauchy surface $\Gamma''$ is causally connected only to the exterior of the black hole, $\Gamma_{ex}$, and hence the unitarity assumption implies that the wavefunctions $\Psi(\Gamma'')$ and $\Psi(\Gamma_{ex})$ are related by a unitary transformation $\mathcal{U'}$,   
	\begin{equation}
	\Psi(\Gamma'') = \mathcal{U'}~\Psi_{ex}(\Gamma_{ex}).
	\end{equation}
	Since we originally started with a pure state $\Psi(\Gamma)$, the two unitary operations imply that $\Psi_{ex}(\Gamma_{ex})$ is also a pure state. This necessitates $\Psi(\Gamma')$ is a product of pure states,
	\begin{equation}
	\Psi(\Gamma') =\Psi_{in}(\Gamma_{in})\otimes\Psi_{ex}(\Gamma_{ex}).
	\end{equation}
	Note that $\Psi_{ex}(\Gamma_{ex})$ is unitarily related to $\Psi(\Gamma)$. If we now require $\Psi(\Gamma)\rightarrow\Psi(\Gamma')$ to be linear, this necessitates that the wavefunction describing the interior of the black hole, $\Psi_{in}(\Gamma_{in})$, cannot have any dependence on $\Psi(\Gamma)$~\cite{complementarity,infopara}. In other words, the quantum state of the interior of the black hole does not contain any information of the quantum state prior to the formation of the black hole; something happens at the event horizon in such a way that it prevents information from entering the black hole. This can be considered as a natural way of requiring a final state boundary condition for the black hole, reconciling unitarity, linearity and the quantum physics of a black hole. In this paper, we adopt the above approach to the black hole information problem, and show that the physics of Andreev reflections in normal metal/ superconducting systems has many of the same features.
	
	There are other proposals which give a special status to the event horizon, and the notion of a \textit{firewall} at the event horizon is one such~\cite{firewall}: the infalling observer encounters a flux of high energy particles (released from breaking of entangled quasiparticle pairs) at the event horizon and gets obliterated. This proposal would similarly contradict the equivalence principle of general relativity according to which the infalling observer should experience an approximately flat space time as he/she crosses the event horizon and nothing special can happen there. In particular, the infalling observer should not encounter a \textit{firewall} at the horizon if we require that the equivalence principle is sacred; no spacetime point can be given a special status as such, which would be a violation of the equivalence principle. Yet another take on this problem is the notion that the black hole is a fuzzball~\cite{mathur2009fuzzballs}; the model assumes reversibility, in that the quantum information encoded in the infalling matter escapes as correlations among the emitted Hawking quanta.     
	
	In this paper we note that the quantum physics of a superconducting condensate share a surprising amount of similarity with the quantum physics of black holes, and the unitary models of black hole evaporation in particular. When a normal metal becomes superconducting, the wavefunction of the superconducting condensate is where all the electrons are paired, and hence exists in a state (the BCS ground state~\cite{BCS}) independent of the initial quantum state of the fermions that formed the condensate. An interesting question to ask now is the following: \textit{what happens to the information falling into the superconductor after the formation of the condensate?} This process is called Andreev reflection~\cite{andreev,Pannet,spintron,artemenko1978excess,artemenko1979theory,artemenko1979excess,zaitsev1980theory}, where electrons incident from the normal metallic region falls into the condensate dragging another electron from the metal, forming a Cooper pair. In the process the information about the incident electron is reflected in the hole that is left behind. To make our analogy precise, we discuss Andreev reflections from the perspective of the Horowitz-Maldacena model, which also necessitates a unique final state for the black hole~\cite{horowitz,projection}. Our black hole analogy based on superconductors has this additional advantage that the pairing mechanism is also included, which is not present in the Bose-Einstein condensate (BEC) models of black holes. The formation of pairs~\cite{horowitz,projection} imply that the infalling Hawking quanta from the horizon also enter the black hole as entangled pairs, which leaves the emitted Hawking radiation in a paired entangled state: the correlations previously existing between the particle--anti-particle pairs at the event horizon are swapped to the correlations between particle pairs and anti-particle pairs. In the special case when a hole pair enters the superconductor (Andreev reflection by a hole), the superconductor releases a Cooper pair into the normal metal. We also note that a large superconducting region may be stable against such \textit{evaporation}: the background lattice will eventually accumulate charge as the Cooper pairs leave the superconducting condensate. A steady state is established for macroscopic superconductors, where the pairs of holes/electrons are transferred between the normal metal and the superconductor at the same rate. Note that the Horowitz-Maldacena model also suggests a similar equilibrium at the event horizon for bigger black holes which do not evaporate~\cite{horowitz}.        
	
	Hayden and Preskill's black hole information mirror model has a similar analogy that can be made~\cite{preskill}: they consider an additional memory system maximally entangled with the infalling matter and a black hole entangled with the Hawking radiation. When the matter falls into the black hole, it becomes maximally entangled with the black hole and that causes the memory system to be maximally entangled with the outgoing Hawking radiation. It is implied that the information has been transferred from the infalling matter to the outgoing Hawking radiation. Both Hayden and Preskill's model and the Horowitz-Maldacena model effectively consider the black hole close to its final state as a mirror which does not take any information, but reflects all the information, while accepting the infalling particles. While naively, this would appear to lose the quantum information, and take the outside pure state to a mixed state, in fact, it can be shown (and will be discussed in detail later in this paper) that the information is in fact reflected from the black hole, leaving the quantum state pure.
	
	We also note that even though the superconductor is able to perform deterministic formation of entangled singlets, there is no \textit{superluminal} transfer of information happening within the superconductor. The entangled pairs in Andreev reflection are local; they are either interacting via the pairing interaction within the superconducting condensate (mediated by phonons) or via the tunneling interaction at the interface. The limiting factor for information transfer here is the pairing interaction within the superconductor; the speed at which information gets transferred through the condensate is roughly equal to the velocity of phonons in the lattice.   
	
	This paper is composed as follows: In Section~\ref{sec2}, we present the dynamics of the spin degree of freedom of electron during Andreev reflection process as a deterministic teleportation. We identify similarities between black hole evaporation models and the analogous Andreev reflection processes in Section~\ref{sec3}. We also discuss the analogy between a traversable Einstein-Rosen bridge~\cite{wormhole} and Crossed Andreev Reflections (CAR)~\cite{car} in Section~\ref{sec4}, and further conjecture that the black hole final state in Horowitz-Maldacena model is same as the BCS wavefunction for the superconductor~\cite{BCS} in Section~\ref{finalstate}. We also relate the thermodynamic properties of a black hole to that of a superconducting condensate.
	\section{Andreev reflection\label{sec2}} 
	\begin{figure}
		\includegraphics[scale=0.4]{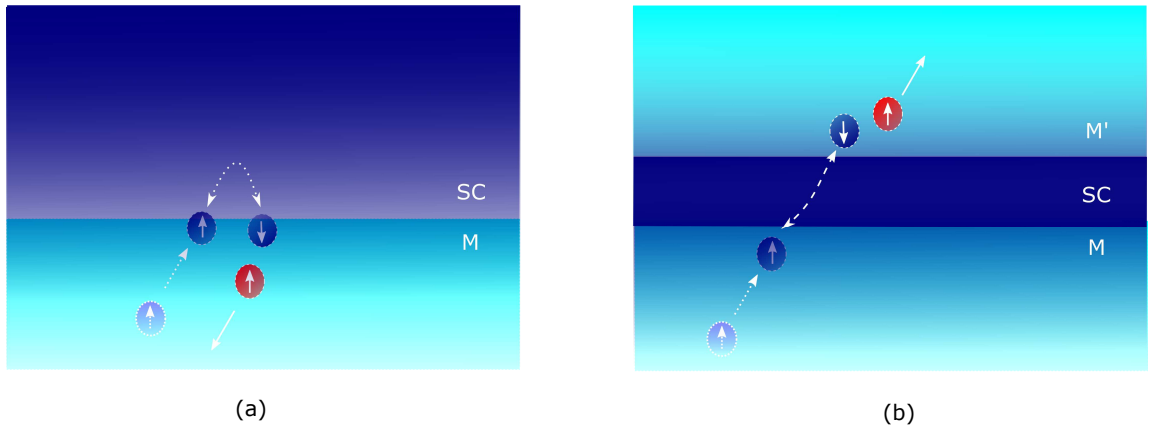}
		\caption{(a) Direct Andreev reflection: an electron (hole) like quasiparticle incident on the superconductor (SC) interface from the metal M, having an energy $\mathcal{E}$ lower than the superconducting energy gap $\Delta$ is retroreflected as a hole (electron) like quasiparticle into the same metal. (b) Crossed Andreev reflection: An electron (hole) like quasiparticle incident from the metal M located at one side of a superconductor -- having a width comparable to the superconducting coherence length -- can be converted to a hole (electron) like quasiparticle in metal M'. The information encoded in the spin degree of freedom is preserved in Andreev reflections.\label{dca}}
	\end{figure} 
	In the conventional approaches, Andreev reflection is studied as a scattering process which occurs when a quasiparticle (electron or hole) from a normal metal -- having an energy $\mathcal{E}$ small compared to the superconducting energy gap $\Delta$ -- is incident on the superconductor~\cite{andreev,Pannet,spintron,artemenko1978excess,artemenko1979theory,artemenko1979excess,zaitsev1980theory}. The incident quasiparticle is retro-reflected as the oppositely charged quasiparticle into the normal metal, with a velocity that is (approximately) opposite to the incident. This results in a supercurrent through the interface transporting a net charge of $-2e$, equivalent to one Cooper pair, across the junction [see Fig.~\ref{dca} (a)].
	Andreev reflections also provide a mechanism to understand how the \textit{proximity effect} operates at a microscopic level~\cite{Pannet,proximity2004}: in Andreev reflections, metal electrons near the superconductor get converted to Cooper pairs, extending the superconducting correlations into the metal.
	
	Here, we consider an electron-like quasiparticle incident from the metal, in the state $|\phi_{e}\rangle$, 
	\begin{equation}|\phi_{e}\rangle =\sum\limits_{k,\sigma}\langle k\sigma|\phi_{e}\rangle e_{k\sigma}^{\dagger}|0\rangle =  (a~e^{\dagger}_{k\uparrow}+b~e^{\dagger}_{k\downarrow})|0\rangle,\end{equation} 
	where $e^{\dagger}_{k\sigma}$ denotes the creation operator for a free electron with wave vector $k$ and spin $\sigma$. The	spin state of the electron in $|\phi_{e}\rangle$ is assumed to be a superposition state of $\uparrow$ spin and $\downarrow$ spin.
	We show that if the superconducting condensate is spherically symmetric (\textit{s wave}), it can accept this spin state by retro-reflecting a hole in the same spin state. The incident quasiparticle mode ($e$) is assumed to have momentum greater than the Fermi momentum ($	\textbf{p}_{e} = \hbar \textbf{k} = \textbf{p}_{F}+\delta \textbf{p}$), and excitation energy $\varepsilon_{e} = \textbf{v}_{F}.\delta\textbf{p}$, relative to the Fermi energy~\cite{spintron}.
	The metal is otherwise described as a filled Fermi sea $|G\rangle$, where all electronic states are occupied up to the Fermi level~\cite{mahan}:
	\begin{equation}
	|G\rangle = \prod_{|k|<k_{F}}~c_{k\uparrow}^{\dagger}c_{k\downarrow}^{\dagger}|0\rangle.\end{equation}
	Here $|0\rangle$ is the particle vacuum.  The resonant interaction between the metal electrons and the electrons in the superconducting condensate at the interface is given by $H_{C}$:
	\begin{equation} 
	H_{C} = \sum\limits_{k,\kappa,\sigma}(j_{k,\kappa}~d_{k,\sigma}^{\dagger}c_{\kappa,\sigma}+\rm{H.c}).\end{equation} 
	Here $j_{k, \kappa}$ is the coupling corresponding to the tunneling interaction, which is Coulombic in nature and hence assumed to be real. The operator $d^{\dagger}_{k\sigma}$ labels the electronic states of superconductor, while $c^{\dagger}_{\kappa\sigma}$ labels the electronic states of the normal metal. Andreev reflection at the interface can be studied by considering only the terms in $H_{C}$ that correspond to $k=\kappa$, as we will assume in the following paragraphs~\cite{ARsecond}. The tunneling Hamiltonian creates spin singlet electron-hole pairs from the Fermi level with one of the particles falling into the superconductor. This can be seen immediately by treating the tunneling interaction perturbatively. The first order term goes like $H_{C}|G\rangle$ which gives us singlet electron-hole pairs that have similar form to the singlet pairs we describe in Appendix~\ref{Ap:excitations}. We take this process to be the analogue of quantum fluctuations near a black hole event horizon: particle-antiparticle pairs are created and annihilated spontaneously, and the Hawking radiation is caused by the antiparticles falling into the black hole, before the pair annihilates. The creation and annihilation of particle-antiparticle pairs in our analogy simply means that the tunneling Hamiltonian creates resonances across the interface where the electron tends to be at either side of the interface. We can also find a stationary state of the Hamiltonian for the interface in this sector, which is a superposition of zero, one, and two excitations:
	\begin{eqnarray}
	\label{interface}	&&\bigg[\sum\limits_{\sigma}\epsilon_{k}(d_{k,\sigma}^{\dagger}d_{k,\sigma}+c_{k,\sigma}^{\dagger}c_{k,\sigma}) +  j(d_{k,\sigma}^{\dagger}c_{k,\sigma}+c_{k,\sigma}^{\dagger}d_{k,\sigma}) \bigg]\bigg[1 + (d_{k,\uparrow}^{\dagger}c_{k,\uparrow}+d_{k,\downarrow}^{\dagger}c_{k,\downarrow})+ d_{k,\uparrow}^{\dagger}c_{k,\uparrow}d_{k,\downarrow}^{\dagger}c_{k,\downarrow}\bigg]c_{k\uparrow}^{\dagger}c_{k\downarrow}^{\dagger}...|0\rangle\nonumber\\
	&&=2~(\epsilon_{k}+j)\bigg[1 + (d_{k,\uparrow}^{\dagger}c_{k,\uparrow}+d_{k,\downarrow}^{\dagger}c_{k,\downarrow})+ d_{k,\uparrow}^{\dagger}c_{k,\uparrow}d_{k,\downarrow}^{\dagger}c_{k,\downarrow}\bigg]c_{k\uparrow}^{\dagger}c_{k\downarrow}^{\dagger}...|0\rangle.
	\end{eqnarray}
	Here $\epsilon_{k}$ is the kinetic energy of the mode $k$, and $j$ is the coupling corresponding to the tunneling interaction.	The second term in the eigenstate describes a singlet excitation, discussed in Appendix~\ref{Ap:excitations}. The last term is a second order excitation, which describes the simultaneous excitation of two quasielectrons, creating two holes in the Fermi sea. In addition to being a second order process in perturbation theory, the simultaneous creation of two holes in the Fermi sea is suppressed by a Boltzmann's factor of $e^{-2\beta\varepsilon_{h}}$ in the steady state at temperature $T$, where $\varepsilon_{h}$ is the excitation energy of a hole, and $\beta = \frac{1}{k_{B}T}$. Here we restrict to first order processes at the interface that create singlet electron-hole pairs. Importantly, all excitations of the superconductor have an energy gap $\Delta$, so any particles incident on the superconductor with energy $\mathcal{E}<\Delta$ must either be reflected or form another Cooper pair in the condensate. We write the total Hamiltonian  for the metal-superconductor junction as a sum:
	\begin{equation}
	H = H_{S}+H_{M}+H_{C} = H_{0}+H_{C}.
	\end{equation}	    
	Here $H_{S}$ is the BCS mean field Hamiltonian for the superconductor~\cite{BCS}:
	\begin{equation}\label{hs} H_{S}=\sum\limits_{k,\sigma} \epsilon_{k}d_{k,\sigma}^{\dagger}d_{k,\sigma}+H_{\Delta},\hspace{1cm}H_{\Delta}=\sum\limits_{k}(\Delta ~d_{k,\uparrow}^{\dagger}d_{-k,\downarrow}^{\dagger}+\text{H.c}),\hspace{1cm}\Delta = -|V_{0}|^{2}\sum\limits_{k^{'}}\langle d_{-k^{'}\downarrow}d_{k^{'}\uparrow}\rangle, \hspace{1cm}\sigma=\uparrow,\downarrow.\end{equation}
	In the mean field description, the superconducting condensate behaves like a scatterer for the electron pairs labeled with momentum $k$ to momentum $k^{'}$ via the mean field interaction Hamiltonian $H_{\Delta}$, though in the mean field approximation, various $k'$ contributions are summed over in the definition of $\Delta$. Here $|V_{0}|$ is the strength of the attractive interaction considered in the BCS model~\cite{annett2004superconductivity}. The metal has energy levels occupied up to the Fermi level: 
	\begin{equation}
	H_{M} = \sum\limits_{\kappa, \sigma}\epsilon_{\kappa}c_{\kappa,\sigma}^{\dagger}c_{\kappa,\sigma}+\sum\limits_{k}U_{M}N_{k\uparrow}N_{k\downarrow}.
	\end{equation}
	The last term in the Hamiltonian $H_{M}$ stands for the short-range Coulomb interaction between the electrons in the metal with number operator $N_{k\sigma}=c_{k\sigma}^{\dagger}c_{k\sigma}$. In order to simplify the tunneling problem, we restrict to the subspace of a single electron incident from the metal, an electron-hole pair at the interface, and a pair of unoccupied electronic states in the superconductor. The incident electron from the metal interacts with the unoccupied electronic states within the superconductor via the tunneling interaction $H^{'}_{C}$:
	\begin{equation}H^{'}_{C} = \sum\limits_{\sigma}[ ~j~d_{k,\sigma}^{\dagger}e_{k,\sigma}+\textrm{H.c.}],
	\hspace{0.5cm}\textrm{such that}\hspace{0.5cm} H^{'}_{C}|\phi_{e}\rangle = j|\phi_{d}\rangle\hspace{0.5cm}\textrm{and}\hspace{0.5cm}H^{'}_{C}|\phi_{d}\rangle = j|\phi_{e}\rangle,\end{equation}
	We assume  that the electronic states localized on the metal are orthogonal to the ones localized on the superconductor, following the treatment in~\cite{ARsecond}. The different modes nevertheless have the same energy, and this also means that $H^{'}_{C}$ commutes with the kinetic term in the Hamiltonian. Hence, up to global phase factors, the state of the incoming mode after the interaction in the Schr\"{o}dinger picture is given by,
	\begin{equation}\label{sincos}
	|\phi(\tau)\rangle=e^{-i\frac{\tau}{\hbar}H^{'}_{C}}|\phi_{e}\rangle = \sin(\frac{j\tau}{\hbar})|\phi_{d}\rangle+\cos(\frac{j\tau}{\hbar})|\phi_{e}\rangle.
	\end{equation}
	The duration of interaction $\tau$ is set by the smallest time required to make a change in the quantum state, bounded from below by the energy time uncertainty principle. Here the uncertainty in energy, $\Delta\epsilon$ is equal to $j$, since a resonant interaction of the form $H^{'}_{C}$ splits the degeneracy of orthogonal states with energy $\epsilon$ into non-degenerate orthogonal states having energies $\epsilon\pm j$. We assume complete resonant transfer of the modes $e^{\dagger}_{k,\sigma}\rightarrow d^{\dagger}_{k,\sigma}$ and hence the choice 
	$\tau \simeq \frac{\pi\hbar}{2~j}$
	is made which also satisfies the energy time uncertainty principle, $j\tau >\frac{\hbar}{2}$. This assumption is made also to be in accord with the conventional treatments of Andreev reflections, where it is not required to distinguish between the incoming mode $e^{\dagger}_{k,\sigma}$ and the infalling mode $d^{\dagger}_{k,\sigma}$; the superconductor is considered as a scatterer, so the microscopic interactions at the interface can be ignored while computing the asymptotic scattering matrix~\cite{andreev}. Note that the requirement for complete resonant transfer of modes can be weakened without affecting the unitarity of the process, as the above resonant interaction at the interface is unitary for all time $\tau$; this in general corresponds to superposition states of the incoming electron and the outgoing hole at the interface. 
	
	Substituting the estimate for $\tau$ into the Eq.~\eqref{sincos}, we obtain $|\phi(\tau)\rangle=|\phi_{d}\rangle.$
	Hence the combined state $|\psi\rangle$ after the interaction at the interface  has the following form:
	\begin{equation}|\psi\rangle = |\phi_{d}\rangle\otimes(d_{q\uparrow}^{\dagger}c_{q\uparrow}+d_{q\downarrow}^{\dagger}c_{q\downarrow})|G\rangle,\end{equation}
	describing an electron incident from the metal and an electron-hole pair at the interface.	Here $q =-k_{F}+\delta k$. This corresponds to a momentum for the excited quasi electron, $\textbf{p}=-\textbf{p}_{F}+\delta \textbf{p}$, and excitation energy $\varepsilon = -\textbf{v}_{F}.\delta \textbf{p} = -\varepsilon_{e}$~\cite{spintron}. From the electron-hole symmetry argument, we can conclude that the hole-like quasiparticle has momentum $\textbf{p}_{h}=-\textbf{p}$ and energy, $\varepsilon_{h} = -\varepsilon = \varepsilon_{e}$~\cite{spintron}. We can rewrite the state $|\psi\rangle$ in a form that allows us to meet the boundary conditions from the superconducting condensate:
	\begin{eqnarray}
	\label{eq:boundary}
	|\psi\rangle &=&\hat{\beta}_{+}|0\rangle(a~c_{q\uparrow}+b~c_{q\downarrow})|G\rangle+\hat{\beta}_{-}|0\rangle(a~c_{q\uparrow}-b~c_{q\downarrow})|G\rangle+\hat{\beta'}_{-}|0\rangle(a~c_{q\downarrow}-b~c_{q\uparrow})|G\rangle+\hat{\beta'}_{+}|0\rangle(a~c_{q\downarrow}+b~c_{q\uparrow})|G\rangle~~\\\nonumber\\
	&=& \hat{\beta}_{+}|0\rangle\otimes i\sigma_{y}|\phi_{h}\rangle + \hat{\beta}_{-}|0\rangle\otimes \sigma_{x}|\phi_{h}\rangle-\hat{\beta'}_{-}|0\rangle\otimes|\phi_{h}\rangle-\hat{\beta'}_{+}|0\rangle\otimes\sigma_{z}|\phi_{h}\rangle.
	\end{eqnarray}
	Here $\sigma_{\alpha},~~\alpha = x,y,z,$ are the familiar Pauli matrices and $\hat{\beta}_{\pm}|0\rangle,\hat{\beta'}_{\pm}|0\rangle$ represent entangled electron pairs within the superconductor, as described in Appendix~\ref{Ap:teleport}.
	\subsection{Applying boundary conditions at the interface}
	We are looking for solutions of the Andreev reflection process with a given energy $\mathcal{E}<\Delta$, which reduces this to a problem of solving time independent Schr\"{o}dinger equation by requiring that all the boundary conditions are met. For this purpose, we identify three regions in the problem: the normal metal, the interface, and bulk of the superconductor. Normal metal allows free propagating electrons and holes as stationary solutions. The stationary solution for the bulk of the superconducting condensate is the BCS ground state wavefunction. We found that the state of incoming modes at the interface can be written as in Eq.~\eqref{eq:boundary}, with the energy-time uncertainty principle taken into account. Now applying the boundary condition is made simple because each term in Eq.~\eqref{eq:boundary} have different orthogonal spin symmetries while the \textit{s wave} superconducting ground state possess only the singlet spin symmetry.
	
	We can also describe the process dynamically, by taking into consideration the pairing interaction within the superconducting condensate. The effective interaction from phonons considered in the BCS theory takes the form of a delta function in the real space~\cite{annett2004superconductivity},
	\begin{equation}
	V = -|V_{0}|^{2}\delta(\textbf{r}_{1}-\textbf{r}_{2}).
	\end{equation}
	Where $|V_{0}|$ is the strength of the attractive interaction.	The electron pair states require nonzero amplitude when $\textbf{r}_{1}=\textbf{r}_{2}$ to feel the attractive interaction. This picks out the singlet spin symmetry for the interior because only the singlet electron pair has a spatial wavefunction which is symmetric, and feel the attractive interaction. More general attractive interactions can be constructed which also permits \textit{s wave} symmetric ground states, such as the potentials in electrostatics and gravity.
	
	We further note that in the mean field approach to Andreev reflections, the superconducting condensate is assumed to act like a source and sink for pairs of electrons with momentum $k$ scattering into electrons with momentum $k'$, where various $k'$ are summed over in the definition of an effective interaction amplitude $\Delta$ as evident from Eq.~\eqref{hs}. The deviation $\delta|\psi'\rangle$ from the initial state $|\psi\rangle$ up to the smallest order in $\Delta$ is,  
	\begin{equation}
	\delta|\psi'\rangle \propto H_{\Delta}|\psi\rangle.
	\end{equation}
	The scattering interaction $H_{\Delta}$ annihilates every other term of $|\psi\rangle$ except the one term with $\beta^{'}_{-}$ due to the singlet symmetry of the Hamiltonian. This is another way of saying that the \textit{s wave} condensate only allows penetration of singlet electron pairs in to the condensate. 
	We can impose this boundary condition by requiring that the bulk of the condensate applies a final state projection onto the superconducting ground state wavefunction~\cite{BCS}, 	\begin{equation}\label{bcsstate}|\Psi_{BCS}\rangle = \prod_{k}(u_{k}+v_{k}e^{-i\chi} d_{k\uparrow}^{\dagger}d_{-k\downarrow}^{\dagger})|0\rangle.\end{equation}
	Partially projecting the quantum state in Eq.~\eqref{eq:boundary} on to this final state in the subspace of the superconductor transfers the quantum information in the incident electron to the hole and also adds the extra phase factor $e^{i\chi}$ which is the superconducting phase of the condensate. We only focus on the non-vanishing term:
	\begin{equation}\langle0|e^{\hat{N}_{\kappa}e^{i\chi}}|\psi\rangle \propto \langle0|e^{\hat{N}_{\kappa}e^{i\chi}}\hat{N}_{\kappa}^{\dagger}|0\rangle\otimes|\phi_{h}\rangle =\langle 0|(1+ e^{i\chi}\hat{N}_{\kappa}+e^{2i\chi}\hat{N}_{\kappa}^{2}/2)\hat{N}_{\kappa}^{\dagger}|0\rangle|\phi_{h}\rangle = e^{i\chi}|\phi_{h}\rangle.\end{equation}
	Here $\hat{N}_{\kappa}^{\dagger}$ is the Cooper pair creation operator we discuss in the Appendices~\ref{Ap:teleport} and ~\ref{Ap:bcscoherent}. In reality,	the electron and hole amplitudes penetrate  a finite distance into the superconductor before the final state projection, and hence the retro-reflected hole also acquires an extra phase factor of $\arccos({\varepsilon/\Delta})$ in Andreev reflections due to this phase delay~\cite{spintron}.	Physically, the incoming electron forms a Cooper pair singlet within the superconductor by taking in a negative energy quasielectron from the Fermi sea, retro-reflecting a positive energy hole. The quasiparticle excitation w.r.t to the Fermi level are formed in the singlet state, and the process of one quasiparticle falling into the superconductor can be thought of as breaking the entangled quasiparticle pair. This ensures that the state of the outgoing hole is  $(a~h^{\dagger}_{-q\uparrow}+b~h^{\dagger}_{-q\downarrow})|0\rangle$, which is exactly the spin state of the incoming electron. This is a quantum teleportation of the spin state of the incoming electron to the outgoing hole, with both the shared entangled pair (the electron-hole pair with momentum $q$) and the final entangled pair (the Cooper pair) always spin singlets! Hence the state of the incoming electron is deterministically transferred to the outgoing hole -- a process called deterministic teleportation~\cite{CTC,projection}. This happens with probability nearly equal to one for ideal metal-superconductor interfaces, and hence we restricted our study only to Andreev reflections~\cite{andreev,Pannet,spintron,artemenko1978excess,artemenko1979theory,artemenko1979excess,zaitsev1980theory}, while ignoring other physical processes due to nonideality of the interface, such as the ordinary reflections of electrons and holes from the interface. The initially shared singlet could be formed on the same metal from which the electron is incident (direct Andreev reflection), or could be on a different metal coupled to the superconductor (crossed Andreev reflection) as shown in Fig.~\ref{dca}.  The above process is particle-hole symmetric, and hence the time-reversal of this process (Andreev reflection of a hole) is also physical. Note that the scattering matrix which relates the asymptotic incoming and outgoing modes in Andreev reflections is unitary, since the information encoded in the incoming mode is same as the information encoded in the outgoing mode.
	\section{Andreev processes analogous to black hole evaporation models\label{sec3}} 
	\begin{figure}
		\includegraphics[scale=0.4]{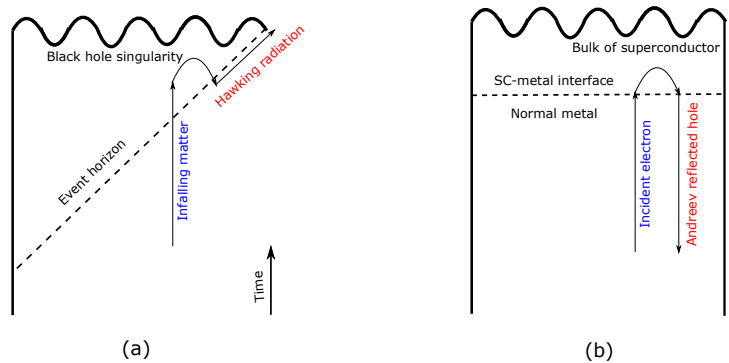}
		\caption{(a) Information dynamics in Horowitz-Maldacena model shown in the Penrose diagram~\cite{penrose2011republication} of a black hole~\cite{projection,horowitz}: The infalling matter takes one Hawking quantum from an entangled particle-antiparticle singlet available at the event horizon. The remaining Hawking quantum escapes to infinity carrying the information. The final state projection on to a singlet inside the horizon ensures that the quantum information has been transfered to the Hawking radiation like quantum teleportation. The arrows indicate that the electron-positron pair at the event horizon teleports the quantm information  encoded in the infalling matter to the exterior of the black hole. (b) Information dynamics in Andreev reflection: An electron incident on the metal-superconductor interface from the metal with an energy $\mathcal{E} <\Delta$, the superconducting energy gap, can be Andreev reflected as a hole in the metal. A Cooper pair singlet is formed in the superconductor. The quantum information encoded in the incident electron is dynamically transferred to the Andreev reflected hole in analogy with the Horowitz-Maldacena final state projection model.\label{andreev}}
	\end{figure} 
	
	In this section, we propose the Andreev processes analogous to two theoretical models for black hole evaporation: the Horowitz-Maldacena mechanism, and Hayden and Preskill's information mirror model. Please see Fig.~\ref{cosmos} and Fig.~\ref{mirror}.       
	We observe that these processes are analogous to a superconductor-induced deterministic teleportation and deterministic entanglement swapping respectively~\cite{projection,niels,bennett1993teleporting,zukowski1993event}, mediated by formation of Cooper pairs in superconducting ground state for a spherically symmetric condensate. Appendix~\ref{Ap:teleport} of this paper gives a brief review of quantum teleportation~\cite{bennett1993teleporting}, and also describes how replacing the Bell measurement device with a particular physical system which can impose a final state boundary condition can be used to achieve deterministic quantum teleportation. It is straightforward to verify from the linearity of the teleportation protocol that if the particle being teleported is entangled with another particle, teleportation preserves the entanglement, via entanglement swapping~\cite{niels,zukowski1993event}.
	\subsection{Andreev reflection and the Horowitz-Maldacena mechanism for black hole evaporation}
	
	The Andreev reflection of a pure spin state at the interface can be thought of as a deterministic teleportation as we described in Sec.~\ref{sec2} and Appendix~\ref{Ap:teleport}.  Here, quasiparticle entangled pairs (excitations of the Fermi sea) available in the junction acts like the initially shared entanglement. They are created due to the proximity with the superconductor, as resonances between the electronic states of the metal and the superconductor at the interface.  The infalling electron forms a Cooper pair within the superconductor by retro-reflecting a hole in the same spin state as the incoming electron. The entire process is unitary since the formation of a singlet pair inside the condensate is not due to a Bell state measurement, but due to a pairing interaction that creates only singlet states. The superconductor can also lose Cooper pairs and shrink in size (evaporate) if the incident quasiparticle is a hole. In BCS theory, holes just below the Fermi sea can form Cooper pairs in the superconducting condensate similar to the electrons just above the Fermi sea~\cite{annett2004superconductivity}. This explains Andreev reflection by a hole where the retro-reflected quasiparticle is an electron.     
	
	This is exactly how the Horowitz-Maldacena model for black hole evaporation works~\cite{horowitz}. Particle/antiparticle pairs are available in the spin-singlet state at the event horizon, and one of them (which has the negative energy) falls into the black hole with the infalling matter, while the remaining quasiparticle (with positive energy) is emitted towards infinity as the Hawking radiation. The final state projection model proposed by Maldacena and Horowitz ensures that the emitted quanta also carry away the quantum information which was previously encoded in the infalling matter. See Fig.~\ref{andreev} for a pictorial representation of the information dynamics in Andreev reflection as a final state projection model.
	\subsection{Superconductor as an information mirror: Hayden and Preskill's proposal for black hole evaporation}
	\begin{figure}
		\includegraphics[scale=0.6]{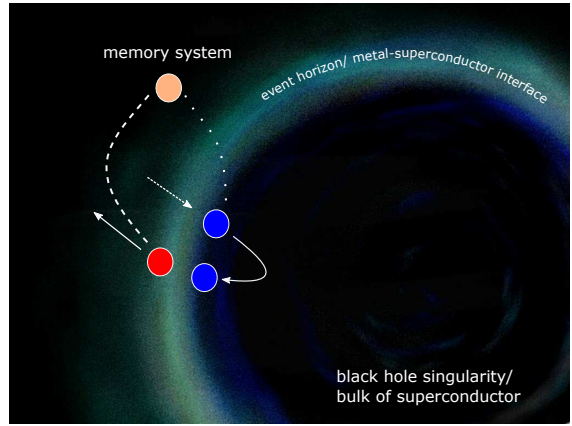}
		\caption{Black hole as an information mirror~\cite{projection,preskill}: Quantum information may exist in form of correlations with an external memory system (Hayden and Preskill's model), and necessitating a final boundary condition for the black hole (bulk of superconductor) will swap the correlations between the infalling particle and the external memory to that between outgoing Hawking radiation (Andreev reflected particle) and the memory system. This process is known as entanglement swapping~\cite{zukowski1993event}.\label{mirror}}
	\end{figure}    
	Hayden and Preskill's mechanism for black hole evaporation assumes a slightly different model for the information entering a black hole~\cite{preskill,projection}: they consider an external memory system maximally entangled with the particle falling into the black hole. The black hole swaps the correlations between the memory system and the infalling particle to correlations between the memory system and the  outgoing Hawking radiation. This can happen in black holes having a final state boundary condition in the following manner~\cite{projection}: Assume the infalling particle ($i$) and the external memory system ($m$) are in a maximally entangled Bell state~\cite{niels}:
	\begin{equation}
	\label{eq:mem}
	|\beta^{m,i}_{00}\rangle =\frac{|0_{m}0_{i}\rangle+|1_{m}1_{i}\rangle}{\sqrt{2}}. 
	\end{equation}
	Any other Bell state can be obtained by local operations on the memory system (or the infalling qubit). The other entanglement we consider is the entanglement at the event horizon, between the two quasiparticles (Hawking quanta, $i$: infalling, $o$: outgoing) spontaneously created out of vacuum:
	\begin{equation}
	|\beta^{H}_{11}\rangle =\frac{|0_{i}1_{o}\rangle-|1_{i}0_{o}\rangle}{\sqrt{2}}. 
	\end{equation}	
	The joint initial state is,
	\begin{equation}
	|\Phi_{in}\rangle = |\beta^{H}_{11}\rangle\otimes|\beta^{m,i}_{00}\rangle.
	\end{equation}
	As the infalling particle descends into the black hole, it also pulls the infalling Hawking quantum along and forms a singlet within the black hole. This is precisely an entanglement swapping~\cite{zukowski1993event}, which causes the memory to be entangled with the outgoing Hawking radiation:
	\begin{equation}
	|\Phi_{f}\rangle = |\beta^{BH}_{11}\rangle\otimes|\beta^{m,o}_{00}\rangle.
	\end{equation}
	In the sense of Hayden and Preskill, this means that the Hawking radiation contains the information now, since the outgoing radiation is maximally entangled with the memory system as in equation~\eqref{eq:mem}. Please refer to Fig.~\ref{mirror} for a pictorial representation of this process.
	
	Now consider an electron falling into the superconductor, which is maximally entangled with an external memory. The other shared entanglement here is between the quasiparticle pairs available at the interface of the normal metal and the superconductor created via the tunneling interaction. The electron falling into the superconductor forms a Cooper pair by absorbing an electron-like quasiparticle from the entangled electron-hole pair, leaving the retro-reflected hole maximally entangled with the external memory. Hence the superconductor behaves like an information mirror similar to the black hole in Hayden and Preskill's description. We predict that such a process does indeed happen in metal superconductor interfaces at equilibrium: the tunneling interaction between the electronic states of the metal and the superconductor creates multiple electron-hole pairs at the interface, while we can also introduce entangled excitations externally using photons. The electron-like quasiparticles may fall into the superconducting condensate and pair up, leaving behind hole pairs in the metal which are correlated, and vice versa. 
	The initial state consists of two electron-hole pairs,  
	\begin{eqnarray}|\psi_{i}\rangle &=& (d_{k\uparrow}^{\dagger}c_{\kappa\uparrow}+d_{k\downarrow}^{\dagger}c_{\kappa\downarrow})(d_{q\uparrow}^{\dagger}c_{q\uparrow}+d_{q\downarrow}^{\dagger}c_{q\downarrow})|G\rangle\\\nonumber\\
	&=&[\hat{\beta}_{+}(c_{q\uparrow}c_{\kappa\uparrow}+c_{q\downarrow}c_{\kappa\downarrow})+\hat{\beta}_{-}(c_{q\uparrow}c_{\kappa\uparrow}-c_{q\downarrow}c_{\kappa\downarrow})+\hat{\beta}^{'}_{+}(c_{q\downarrow}c_{\kappa\uparrow}+c_{q\uparrow}c_{\kappa\downarrow})+\hat{\beta}^{'}_{-}(c_{q\downarrow}c_{\kappa\uparrow}-c_{q\uparrow}c_{\kappa\downarrow})]|G\rangle,\end{eqnarray}
	where $\kappa$ could be potentially different from $k$ in the infalling particle/ memory system depending on the process that created the exciton pair. 	A spherically symmetric superconductor necessitates the final state projection on to the state $e^{\hat{\beta}^{'}_{-}e^{-i\chi}}|0\rangle$, leaving the quantum state in the metal, $
	|\psi_{f}\rangle\propto(c_{q\downarrow}c_{\kappa\uparrow}-c_{q\uparrow}c_{\kappa\downarrow})|G\rangle$, which describes a pair of correlated holes in the normal metal.	
	\section{Mode conversion in Crossed Andreev Reflections (CAR): Superconductor as an Einstein-Rosen bridge\label{sec4}}
	\begin{figure}
		\includegraphics[scale=0.6]{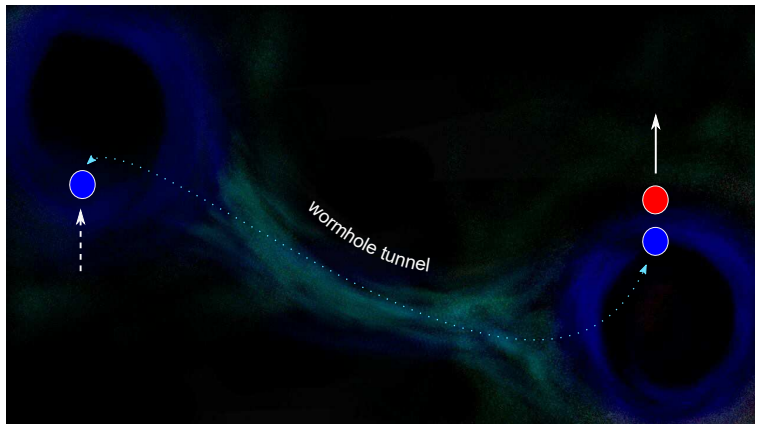}
		\caption{A traversable Einstein-Rosen bridge (a wormhole): electron entering the wormhole (blue dot) is non-locally Andreev reflected as holes in to another universe (red dot). The quantum information is preserved, since the hole carries the same spin state as the original electron.\label{cosmos2}}
	\end{figure}
	
	An interesting microscopic description for an Einstein-Rosen bridge~\cite{einstein1935particle} (wormhole, the name given by Wheeler) is provided by Maldacena and Susskind where they treat the wormhole as a system whose microstates exist as entangled pairs of microstates of two black holes~\cite{wepr}. Here we argue that this picture of a wormhole is quite similar to a superconductor sandwiched between two metals, where metal one and metal two correspond to the two different regions of spacetime external to the wormhole mouths separated by two distinct event horizons. In our analogy, Cooper pairs in the superconducting condensate are the analogue of the microstates of a wormhole in the model described by Maldacena and Susskind~\cite{wepr}.
	
\begin{figure}
		\includegraphics[scale=0.4]{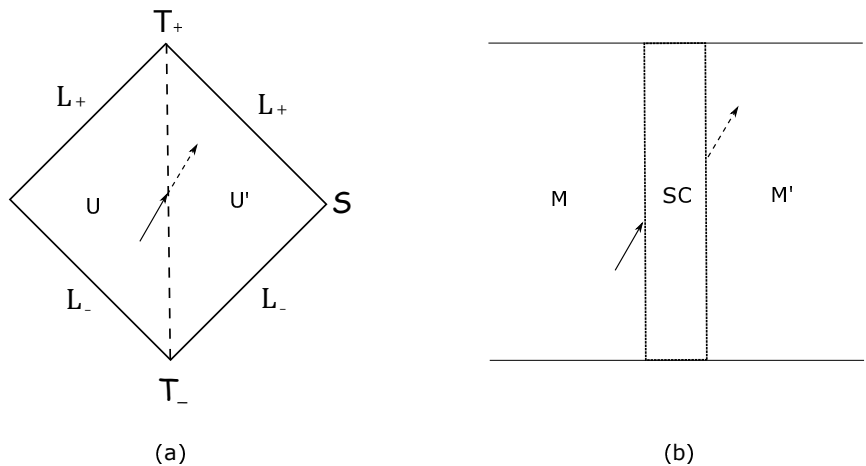}
		\caption{(a) Penrose diagram~\cite{penrose2011republication} of a traversable Einstein-Rosen bridge: World lines starting in the universe U can end up in universe U'. The dashed vertical line is the throat of Einstein-Rosen bridge. We use $\text{L}_{+(-)}$ to represent the future (past) light-like infinity and $\text{T}_{+(-)}$ to represent future (past) time-like infinity. $\mathcal{S}$ represents space-like infinity~\cite{penrose}. (b) Information dynamics in Crossed Andreev Reflection: An electron incident on the metal-superconductor interface from the metal M with an energy $\mathcal{E} <\Delta$: the superconducting energy gap, can be Andreev reflected as a hole in metal M', if the width of the superconducting region is smaller than the superconducting coherence length. The intermediate superconductor can be thought of as a traversable Einstein-Rosen bridge for quantum information encoded in electron spins, transferred between the two metals.\label{car}}
	\end{figure}

	Our analogy for the interior of a black hole to the superconducting condensate also provides a mechanism for wormhole travel of information via \textit{crossed Andreev reflections}.  Similar to the ``direct'' Andreev reflections we discussed in the previous section, Andreev reflections can happen across a superconductor in metal-superconductor-metal junctions where the incidence and retro-reflection could potentially happen at different interfaces. This process, known as Crossed Andreev Reflection (CAR)~\cite{car} has been studied extensively. A mathematical description of CAR in our model would be identical to the one we provide in Sec.~\ref{sec2}, where the electron-hole pair exists at the interface between the superconductor and metal two, for electrons incident from metal one (see Fig.~\ref{dca}). The formation of singlet quasiparticle excitations acts like the shared singlet at the wormhole exit and the formation of a Cooper pair in the superconducting bulk is the final state projection to a singlet pair. The width of the superconducting layer should be of the order of the superconducting coherence length for CAR to occur~\cite{car}, which would need to extend over the entire wormhole for this picture to apply in the gravitational case. The information transfer via the superconducting condensate is mediated by the formation of Cooper pairs, and hence the speed at which information transfer occurs through the condensate can be roughly estimated to be the velocity of phonons in the medium. Information transfer through the superconducting condensate is thus distinctively different from conduction through a normal metal, as the former involves a supercurrent. Please see the analogy depicted in figures~\ref{cosmos2} and~\ref{car}.
	
	Our picture of wormhole travel is quite different from the conventional picturing of wormhole travel as traveling smoothly through a tunnel connecting two different valleys.  In this accounting, the traveler, upon entering the black hole, will collide with the superfluid and immediately have all particles absorbed into the condensate formed by the extreme gravitational attraction.  However, rather than be obliterated, the information contained in the particles of the traveler will be teleported to the other end of the wormhole, via the crossed Andreev reflection process, and be shot out as anti-particles.    
	
	We stress that like in Andreev reflection for metal-superconducting interfaces, there is no necessity for all the information to be teleported across the wormhole.  In general, there will be a combination of regular Andreev reflection and crossed Andreev reflection processed occurring, unless special circumstances are taken into account.  For example, in the N-S interface, crossed Andreev reflection can be promoted over normal Andreev reflection if the metal on one side is also a ferromagnet~\cite{cadden2007charge}.
	
	\section{The final quantum state of a black hole\label{finalstate}}
	The BCS theory for superconductivity is formulated based on the observation that electrons above the Fermi sea form a bound state when subjected to even weak attractive interactions~\cite{annett2004superconductivity}. The attractive interaction in superconductivity is mediated by exchange of phonons which are quantized vibrations of the lattice. In this section, we conjecture that the quantum state of a black hole is achieved similarly, where the attractive interaction is dominated by the force of gravity. A black hole can be formed when the gravitational force exceeds the fermionic degeneracy pressure of the identical fermions that remain in a star, which is proceeding towards the final stages of nuclear fusion forming heavier elements.  When the star is not massive enough to gravitationally collapsing on itself prevailing the fermionic degeneracy pressure, the final state is called a neutron star. The idea that the strong attractive forces within a neutron star can give rise to superfluidity was proposed in the earlier days of superconductivity, even before the first experimental detection of a neutron star~\cite{neutron1}. The analogue of an \textit{s wave} superconductor would be an isotropic neutron star, and they are known to form singlet Cooper pairs~\cite{superfluidity,stellar}. Different kind of pairing is also possible, such as the spin triplet pairing at higher densities~\cite{stellar}.  Recent studies based on observation of cooling of a neutron star also suggest that superfluidity from a BCS-like pairing of neutrons could play a crucial role in explaining the star's rapid cooling~\cite{neutron2}. The fact that the core of a neutron star is thought to be a superfluid gives further support of this idea that black holes are perfect superconductors (when charged) or superfluids, where all fermions are paired. We conjecture that the final quantum state of fermions that form a spherically symmetric black hole is same as the ground state of the superconductor proposed by Bardeen, Cooper and Schrieffer (BCS) given in Eq.~\eqref{bcsstate}~\cite{BCS}.
	
	We assume that all fermions are paired up, and the coherence length of the black hole is the entire Schwarzschild radius. This state satisfies all the symmetry requirements and also explains why black hole evaporation is unitary. Notice that it also avoids sign ambiguities for fermions since they appear as pairs of two. In the cores of neutron stars, the mechanism of pairing is attributed to the strong nuclear forces as the analogue of phonons in BCS theory~\cite{superfluidity,neutron2,neutron1}. Our conjecture is that gravity could play a key role in the pairing mechanism within the black holes: The microstates of a spherically symmetric black hole exists as entangled pairs, mediated by gravity rather than by phonons as in conventional superconductivity.  The theoretical appeal of this conjecture is that (1) it solves the black hole information problem, (2) it gives a specific set of predictions of how particle and antiparticles enter into black holes, (3) it maps existing proposals of how to preserve information entering black holes onto well understood physics in a different context where quantitative experiments can be performed under controlled conditions, and (4) it gives a mechanism for wormhole transfer of information with apparent loss of information in our world.
	
	Further, it is known that extremal black holes (black holes with the smallest possible mass for a given charge and angular momentum) expel electric and magnetic fields similar to a superconductor, and the phenomena is called the \textit{Meissner effect} in extremal black holes~\cite{king1975black}. Geometrically, the throat lengths of extremal black holes tends to infinity and hence the electromagnetic fields originating from sources outside the horizon  decay before they make it to the horizon. A more contextual explanation for the Meissner effect in black holes relates the phenomena to entanglement~\cite{penna2014black}: Robert F. Penna has shown that the two-point correlation function across the horizon in the Hartle-Hawking vacuum vanishes as temperature is lowered, which is interpreted as the vanishing of the entanglement between modes on either side of the event horizon. We note that a similar vanishing of entanglement is also true for the final state projection models we discussed in this paper: the interior quantum state of a black hole is pure in the final state projection models, and this final state is achieved by breaking the entanglement between modes across the event horizon~\cite{projection,horowitz}.
	
	The particular final state boundary condition  we choose for the interior of a black hole, which is the BCS ground state wavefunction, also predicts the Meissner effect for charged fermionic condensates~\cite{BCS}. Meissner effect in superconductors can be derived as a property of the BCS ground state, from its off-diagonal long range order~\cite{rensink1967off,nieh1995off}. In systems that maintain off-diagonal long range order, the features of many body quantum state is extractable from the eigenstate of the reduced density matrix for a single pair corresponding to the largest eigenvalue, and in superconductivity, it can be related to the Ginzburg-Landau
	wavefunction~\cite{nieh1995off}. This would suggest that the charged condensates at the interior of extremal black holes can also contribute to the Meissner effect just like a superconductor, for example, the extremal charged black hole in Reissner-Nordstrom metric~\cite{bivcak1980stationary,penna2014black}. 
	
	The unique quantum wavefunction for the final state of black holes that we conjecture here is limited in the sense that it considers fermions only. Gravitons are naively included in our discussion as particles mediating the pairing interaction, analogous to phonons in BCS theory. While this consideration alone is interesting and shares a surprising amount of similarity to the quantum physics of a black hole, we emphasize that a complete final state description of black holes should prescribe final states for all kind of particles, including bosons.    
	\subsection{The entropy and temperature}   
	The proposed BCS ground state wavefunction for the black hole has interesting thermodynamic properties analogous to a black hole. It was shown by Puspus et al., that the entanglement entropy of the BCS ground state wavefunction scales like area~\cite{puspus2014entanglement}. We show that this area scaling of entropy is closely related to Andreev reflections happening at the interface between a normal metal and the superconductor: the pairs that contribute the largest to the entropy are exactly the those pairs enter/leave the condensate during Andreev reflections. We also compare the temperature of the BCS ground state to the temperature of a black hole.
	
	Area law for entanglement entropy of the BCS ground state~\cite{puspus2014entanglement}: In the BCS state, Cooper pairs are formed for all $k$ vectors in the Debye shell $\varepsilon_{k}~\epsilon~[\varepsilon_{F}-\varepsilon_{D},\varepsilon_{F}+\varepsilon_{D}],$ where $\varepsilon_{D}$ is the Debye energy of the lattice. Puspus et al., have shown that when the pairing energy $\Delta$ is small compared to the Fermi energy, $\varepsilon_{F}$, the major contribution to the entanglement entropy of the BCS ground state comes from orbitals with $\varepsilon_{k}\simeq\varepsilon_{F}$. The entanglement entropy in this case is proportional to the number of states on the Fermi surface, which scales like an area~\cite{puspus2014entanglement}.  We notice that Andreev reflections provide an interesting new perspective of looking at this result: Andreev reflections from the interface can potentially create the pairs that contribute the largest to the entanglement entropy of the bulk of the superconducting condensate, since the pairs which enter/leave the superconductor during Andreev reflections have momentum closely equal to the Fermi momentum~~\cite{andreev,Pannet,spintron,artemenko1978excess,artemenko1979theory,artemenko1979excess,zaitsev1980theory}. In other words, the information contained in the BCS ground state wavefunction, measured by the entanglement entropy, correponds to a physical process (Andreev reflections) at the boundary which can be thought as the microscopic origin of this entropy. This observation further supports our analogy of Andreev reflections to Hawking radiation in black holes, in that the latter is conjectured to be the microscopic origin of entropy of black holes~\cite{srednicki1993entropy,das2008black,bombelli1986quantum}.
	
	Temperature of the BCS ground state~\cite{puspus2014entanglement}: Puspus et al., computes the entanglement entropy of the BCS ground state from the reduced density matrix, obtained by tracing out the spin-down electrons. They have also noted that this reduced density matrix for spin-up electrons can be approximated to a canonical Gibbs ensemble with a constant inverse temperature~\cite{puspus2014entanglement}:
	\begin{equation}
	\label{Tge}
	\frac{1}{T_{Ge}} =\frac{2~k_{B}}{\Delta}\coth^{-1}\sqrt{2}\simeq\frac{1.7627~k_{B}}{\Delta}.
	\end{equation}
	The canonical Gibbs ensemble and the associated temperature $T_{Ge}$ can be interpreted as the effective thermodynamic description of the BCS ground state. 
	Further $T_{Ge}$ is approximately equal to the inverse critical temperature for the superconductor, $\frac{1}{T_{C}} = \frac{\pi e^{-\gamma}~k_{B}}{\Delta} = \frac{1.7639~k_{B}}{\Delta}$, where $\gamma$ is the \textit{Euler-Mascheroni constant}~\cite{puspus2014entanglement}.  We can rewrite this temperature of the Gibbs ensemble, $T_{Ge}$ ($\simeq T_{C}$), in terms of the coherence length of the superconductor $\lambda$~\cite{annett2004superconductivity}, 
	\begin{equation}
	\lambda = \frac{\hbar v_{F}}{\pi\Delta},
	\end{equation}
	where $\Delta$ is the superconducting energy gap and $v_{F}$ is the Fermi velocity. The superconducting coherence length $\lambda$ is the length over which two electrons maintains the coherence, and it also characterizes the typical size of a Cooper pair bound state in BCS theory~\cite{annett2004superconductivity}. In terms of $\lambda$,   
	\begin{equation}
	T_{Ge} = \frac{1}{k_{B}~\coth^{-1}\sqrt{2}}~\frac{\Delta}{2} = \frac{1}{k_{B}~\coth^{-1}\sqrt{2}}~\frac{\hbar v_{F}}{2\pi\lambda}\simeq 1.1346~ \frac{\hbar v_{F}}{2\pi\lambda~k_{B}}, 
	\end{equation} which is approximately equal to the superconductor's critical temperature,
	$T_{C} \simeq 1.1338~ \frac{\hbar v_{F}}{2\pi\lambda~k_{B}}.$ We are now able to make our analogy more precise by comparing this with the expression for the temperature of a Schwarzschild black hole~\cite{temp,temp2,temp3}, 
	\begin{equation}\label{bhT}T_{BH} =\frac{\hbar c}{4\pi r_{s}~k_{B}},\end{equation}
	where we notice that the Fermi velocity is analogous to the speed of light c, and half the coherence length $\lambda$ is the analogue of the Schwarzschild radius, $r_{s}$ for the black hole. We consider this mapping between the relevant parameters of a black hole and a superconductor by comparing the expressions for their respective temperatures as one of the important results of this paper, in that it relates to the collective behavior of the microstates of the system in the thermodynamic limit, as opposed to  the microscopic quantum physics of Andreev reflections we discussed so far in our analogy.     
	\subsection{Negative specific heat of black holes: a superconducting ground state perspective}
	\begin{figure}
		\includegraphics[scale=0.5]{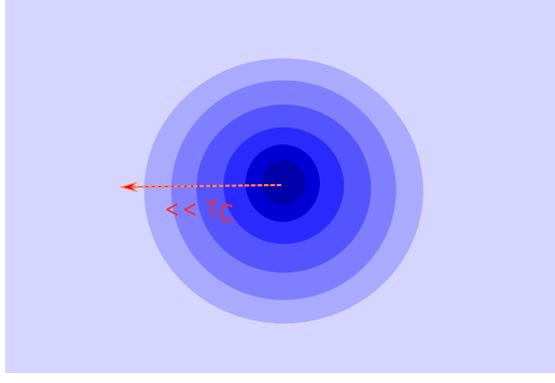}  		\caption{Different \textit{s wave} (spherically symmetric, spin singlet) superconducting domains arranged in the decreasing order of critical temperature $T_{C}$. The size (or mass) of the condensate increases [decreases] as we gradually cool [heat] the system. We conclude that the superconducting condensate, considered as an isolated system, exhibits negative (mass) specific heat.\label{scT}} 
	\end{figure}
	A black hole in an asymptotically flat spacetime has a negative specific heat. When mass (equivalent of energy) is added to the black hole, its temperature decreases. Our analogy suggests that the negative specific heat can be understood as a purely local effect arising from the particular final state of the black hole. To describe how a superconducting condensate wavefunction can demonstrate an equivalent of a negative specific heat, we consider the analogous process of increasing the mass (number of Cooper pairs) of a condensate.  Let us consider an experiment where the critical temperature decreases with the increasing radius of a sphere. This can be done with concentric spheres of materials having critical temperatures $T_{C_{i}}$ that are arranged in the decreasing order of critical temperatures,  as depicted in Fig.~\ref{scT}. Note that our arrangement is also in agreement with the inverse scaling of temperature with the radius of the black hole as in Eq.~\eqref{bhT}. As we cool this system down, the inner domains become superconducting first, followed by the outer domains in the order of their respective critical temperatures. We will see a gradual increase in the number of Cooper pairs in the superconducting condensate.  Since the number of Cooper pairs is a good estimate for the mass of the condensate, $M$, (analogous to the mass of a black hole) we draw the conclusion that,
	\begin{equation}
	C_{M} = \frac{dM}{dT}<0,
	\end{equation}
	for the condensate. To be more precise, when we consider the superconducting condensate as an isolated closed system and ignore the environment completely, the condensate does indeed have a negative (mass) specific heat. Put another way, injecting free electrons into the superconductor will form more Cooper pairs, effectively decreasing the temperature of the superconductor as the mass of the condensate (number of Cooper pairs) increases. 
	
	We now consider a relatively large condensate in proximity to a normal metal. We can make the condensate stable against Andreev reflections by introducing  ordinary reflectors as \textit{walls} at the asymptotic regions of the normal metal. The asymptotic \textit{walls} prevent Andreev reflected particles from escaping to infinity, and the superconducting condensate exists in a canonical equilibrium with Andreev reflected quasiparticles, where electron pairs get added to, and removed from the condensate at the same rate. We speculate that this can happen quite naturally in experiments, as the neighboring metal electrodes have a fixed temperature and chemical potential, and it is also possible to electrically isolate the system by surrounding it with insulators. Our idea is quite similar to Hawking's original idea, which was to place the Schwartzschild black hole in a box in order to achieve equilibrium with the emitted radiation~\cite{hawking1976black}. A black hole in anti-de Sitter (AdS) spacetime is a more physical realization of the proposal by Hawking~\cite{AdS}; the gravitational potential tends to infinity in the asymptotes of AdS spacetime, which prevents Hawking radiation from escaping to infinity. Hawking radiation is emitted and re-absorbed at the same rate, and the black hole exists in canonical equilibrium with the emitted radiation with a  positive specific heat.     
	\section{Conclusions}
	This paper has given an analogy of the quantum physics of superconductor/metal interfaces to that of black holes, while there are obvious differences.  We have proposed the final quantum state of a black hole to be simply the BCS ground state, where all fermions are paired and the coherence length is the Schwarzschild radius.  To support this thesis, we have pointed out that several existing proposals to preserve quantum information entering black holes can be directly mapped to the case of the spin information of electrons in a metal entering the superconductor, with the energy of the electron less than the superconducting gap.  
	
	Using this analogy, we have made predictions for quantum information in superconductors and in black holes.  In the superconducting case, we have shown that quantum information cannot enter it, and that the Hayden and Preskill~\cite{preskill}, and Horowitz-Maldacena~\cite{horowitz} information mirror mechanisms are applicable~\cite{preskill,horowitz,projection}:   Since the superconducting ground state consists of paired electrons in singlet states, while the particle is permitted to enter the superconductor, its spin information is reflected via an Andreev reflection. This process can be viewed as a deterministic quantum  teleportation of information, described by a scattering matrix which is unitary~\cite{projection,horowitz}.  We have made several predictions of quantum gravity experimentally accessible by mapping them to Andreev reflection processes. The final state projection on the BCS ground state resolves the information paradox, while being able to directly apply the Horowitz and Maldacena methods of information preservation via teleportation~\cite{horowitz}.
	Our proposal also gives a precise mechanism of {\it apparent} loss of quantum information in a black hole by the process of nonlocal Andreev reflection, transferring the quantum information through a wormhole and into another universe.  In this process, the entering particles are absorbed into the condensate, while their information is teleported across the wormhole by the black hole coherently drawing in the appropriately spin and momentum paired particle from the other universe and ejecting the information non-locally with positrons from its event horizon.  We have also shown how the equivalent of black hole evaporation happens by the loss of a Cooper pair in the Andreev reflection of a hole, resulting in the shrinking of the superconductor. 
	
	We show that the area law for entanglement entropy of the superconducting ground state is closely related to Andreev reflections. We also make the connections between the temperature of the black holes and the critical temperature of the superconducting condensate, and relate the Fermi velocity to the speed of light and the coherence length of the superconductor analogous to the Schwarzschild radius. The negative specific heat of the black hole is explained in the context of a growing/evaporating condensate, considered as an isolated closed system. All of these phenomena point to our conclusion that the final state of a black hole is just the BCS ground state.  We should stress that the mechanism of the pairing is quite different - in the metallic superconductors, it is the interactions of the electrons with the lattice that leads to the effective attractive interaction.  Here, it is the gravitational force between particles in a curved space time that leads the pairing mechanism to have a lower energy. We note that while the BCS paired ground state is our conjectured final state, the system is certainly strongly interacting, and will likely go beyond the theory of weakly interacting fermions by Bardeen, Cooper and Schrieffer (BCS)~\cite{BCS}.
	\section*{Acknowledgments}We thank Prof. S. G. Rajeev, and Prof. E. Blackman for discussions and suggestions.  This work was supported by the John Templeton Foundation, grant ID 58558.
	\appendix
	\section{Excitations of the Fermi sea\label{Ap:excitations}}
	The Fermi sea is defined by~\cite{mahan}:
	\begin{equation}
	|G\rangle = \prod_{|k|<k_{F}}~c_{k\uparrow}^{\dagger}c_{k\downarrow}^{\dagger}|0\rangle.\end{equation}
	Here $|0\rangle$ is the particle vacuum.
	Excitations about the filled Fermi sea represent electron-like and hole-like quasiparticles. An excitation could be thought of as a promotion of an electron from a band $k<k_{F}$ to a level $k^{'}>k_{F}$ leaving behind a vacancy (hole) in the Fermi sea. The electron originally had a momentum $\hbar k$. By promoting it to an excited state, the total momentum of the Fermi sea changes by a factor $-\hbar k$. This is the momentum associated to the hole. A similar argument can be used to identify the spin of holes w.r.t this redefined ground state. Consider a total spin $\mathcal{S}_{z}$ of electrons in the Fermi sea along the $z$ direction. If the electron being promoted to an excited state had a spin $\sigma$, the total spin along $z$ in the Fermi sea changes by $-\sigma$, which can be thought as the spin of the hole. This is manifested in the particle-hole symmetry relation $h^{\dagger}_{k,\sigma} = e_{-k,-\sigma}$. 
	
	We are interested in the kind of excitations $c_{k+q,\sigma}^{\dagger}c_{k,\sigma}|G\rangle$ where summation over the spins $\sigma$ is implied. Since we work in a picture with a filled Fermi sea as equivalent to a redefined vacuum, we can assume that the absence of quasiparticle excitations correspond to a vacuum with spin zero. We would hence expect, from the conservation of spin, that the quasiparticle excitation we considered is also a spin singlet. To see this explicitly, we rewrite 
	\begin{eqnarray}
	\label{exc}	c_{k+q,\sigma}^{\dagger}c_{k,\sigma}|G\rangle &=& (c_{k+q,\uparrow}^{\dagger}c_{k,\uparrow} +  c_{k+q,\downarrow}^{\dagger}c_{k,\downarrow})|G\rangle = (c_{k+q,\uparrow}^{\dagger}c_{k,\uparrow} +  c_{k+q,\downarrow}^{\dagger}c_{k,\downarrow})\prod_{|k|<k_{F}}c_{k,\uparrow}^{\dagger}c_{k,\downarrow}^{\dagger}|0\rangle\\
	&=&[c_{k+q,\uparrow}^{\dagger}(1-c_{k,\uparrow}^{\dagger}c_{k,\uparrow})c_{k,\downarrow}^{\dagger}|0\rangle +  c_{k+q,\downarrow}^{\dagger}c_{k,\downarrow}c_{k,\uparrow}^{\dagger}c_{k,\downarrow}^{\dagger}|0\rangle]\otimes\prod^{\kappa\neq k}_{|\kappa|<k_{F}}c_{\kappa,\uparrow}^{\dagger}c_{\kappa,\downarrow}^{\dagger}|0\rangle  \\
	&=&[c_{k+q,\uparrow}^{\dagger}(1-c_{k,\uparrow}^{\dagger}c_{k,\uparrow})c_{k,\downarrow}^{\dagger}|0\rangle -  c_{k+q,\downarrow}^{\dagger}c_{k,\uparrow}^{\dagger}(1-c_{k,\downarrow}^{\dagger}c_{k,\downarrow})|0\rangle]\otimes\prod^{\kappa\neq k}_{|\kappa|<k_{F}}c_{\kappa,\uparrow}^{\dagger}c_{\kappa,\downarrow}^{\dagger}|0\rangle\\
	&=&(c_{k+q,\uparrow}^{\dagger}c_{k,\downarrow}^{\dagger} -  c_{k+q,\downarrow}^{\dagger}c_{k,\uparrow}^{\dagger})|0\rangle\otimes\prod^{\kappa\neq k}_{|\kappa|<k_{F}}c_{\kappa,\uparrow}^{\dagger}c_{\kappa,\downarrow}^{\dagger}|0\rangle,
	\end{eqnarray}	
	which is clearly a spin singlet state as expected. We can also write this state in the first quantized form to see this:
	\begin{equation}
	\Psi(r_{1},r_{2}) = \langle r_{1},r_{2}|(c_{k+q,\uparrow}^{\dagger}c_{k,\downarrow}^{\dagger} -  c_{k+q,\downarrow}^{\dagger}c_{k,\uparrow}^{\dagger})|0\rangle = (e^{i(k+q)r_{1}}e^{ikr_{2}}+e^{ikr_{1}}e^{i(k+q)r_{2}})\bigg[\frac{|\uparrow\downarrow\rangle-|\downarrow\uparrow\rangle}{\sqrt{2}}\bigg].
	\end{equation}	
	The missing of an electron from the Fermi sea is interpreted as a hole as we discussed previously. 	The singlet exciton state (here refers to a bound state of an electron and a hole with total spin zero) is rotationally invariant. It hence preserves the electron-hole correlations and thus better explains why Andreev reflections in normal metals do not have any preference for a spin direction.
	\section{Teleportation of quantum information\label{Ap:teleport}}
	Quantum teleportation uses shared entangled pairs and the ability to communicate classically, in order to achieve transfer of quantum information between two locations $A$ and $B$~\cite{bennett1993teleporting}. The locations $A$ and $B$ are equivalently described as connected via a quantum channel (also known as an Einstein-Podolsky-Rosen bridge/pair) which is the shared entangled pair. We consider an entangled singlet $|\beta^{AB}_{11}\rangle$ to be the initially shared entangled pair,
	\begin{equation}
	|\beta^{AB}_{11}\rangle = \frac{|\uparrow\downarrow\rangle-|\downarrow\uparrow\rangle}{\sqrt{2}}.
	\end{equation}
	Any other maximally entangled state would work in a similar way. Consider an arbitrary pure state $|\psi^{C}\rangle$ which is to be teleported from $A\rightarrow B$, 
	\begin{equation}|\psi^{C}\rangle = a|\uparrow\rangle + b|\downarrow\rangle.\end{equation}
	The advantage of quantum teleportation is that no information about the parameters $a$ and $b$ is required to transfer the quantum state. The joint initial state of the three particles is,
	\begin{equation}
	|\psi\rangle = |\psi^{C}\rangle\otimes|\beta^{AB}_{11}\rangle,
	\end{equation}
	which can be written in the Bell basis of particles $C$ and $A$ as:
	\begin{equation}\label{tp}|\psi\rangle = (-|\beta^{CA}_{11}\rangle|\psi^{B}\rangle-|\beta^{CA}_{10}\rangle\sigma_{z}|\psi^{B}\rangle+|\beta^{CA}_{01}\rangle\sigma_{x}|\psi^{B}\rangle+i|\beta^{CA}_{00}\rangle\sigma_{y}|\psi^{B}\rangle)/2,\end{equation}    
	where \begin{equation}|\beta_{ab}\rangle = \frac{|0a\rangle+(-1)^{b}|1\bar{a}\rangle}{\sqrt{2}},\hspace{1cm}\hspace{2cm}|0\rangle = |\uparrow\rangle,\hspace{1cm}|1\rangle = |\downarrow\rangle.\end{equation}
	A Bell basis measurement on the two particles $C$ and $A$ now collapses this state in to one of the four possibilities and in conventional quantum teleportation, the protocol is completed by classically communicating the measurement result from $A\rightarrow B$ and applying an appropriate unitary to recover the original state.  It is interesting to note that statistically a quarter of the times the Bell measurement gives the outcome $|\beta_{11}^{CA}\rangle$ in which case no classical communication is required. In other words, we can achieve teleportation deterministically a quarter of the times if we throw away the other Bell measurement outcomes and postselect on the singlet state.   
	
	The situation is entirely different if there is some way we can impose that the measuring device always gives the singlet state as the measurement outcome. It was proposed by Horowitz and Maldacena that if the Bell measurement device is replaced by a physical system (a black hole in their conjecture), the symmetry of the black hole state can impose a particular pairing (singlet) and hence achieve deterministic quantum teleportation without having to throw away undesired measurement outcomes.
	
	We note that the situation in Andreev reflection is quite similar, where an \textit{s wave} superconductor can impose a singlet pairing to the incoming modes. Note that the state $|\psi\rangle$ in Eq.~\eqref{eq:boundary} has a similar form to that in Eq.~\eqref{tp}:
	\begin{eqnarray}
	|\psi\rangle &=&\hat{\beta}_{+}|0\rangle(a~c_{q\uparrow}+b~c_{q\downarrow})|G\rangle+\hat{\beta}_{-}|0\rangle(a~c_{q\uparrow}-b~c_{q\downarrow})|G\rangle+\hat{\beta'}_{-}|0\rangle(a~c_{q\downarrow}-b~c_{q\uparrow})|G\rangle+\hat{\beta'}_{+}|0\rangle(a~c_{q\downarrow}+b~c_{q\uparrow})|G\rangle\nonumber\\\nonumber\\
	&=& \hat{\beta}_{+}|0\rangle\otimes i\sigma_{y}|\phi_{h}\rangle + \hat{\beta}_{-}|0\rangle\otimes \sigma_{x}|\phi_{h}\rangle-\hat{\beta'}_{-}|0\rangle\otimes|\phi_{h}\rangle-\hat{\beta'}_{+}|0\rangle\otimes\sigma_{z}|\phi_{h}\rangle.
	\end{eqnarray}
	We have used the definition of the Fermi level $|G\rangle:$
	\begin{equation}|G\rangle = \prod_{|k|<k_{F}}~c_{k\uparrow}^{\dagger}c_{k\downarrow}^{\dagger}|0\rangle = |1_{q\uparrow}1_{q\downarrow}...\rangle,\end{equation}
	and the following relations:
	\begin{eqnarray}
	(a~c_{q\uparrow}+b~c_{q\downarrow})|G\rangle &=& a|0_{q\uparrow}1_{q\downarrow}...\rangle-b|1_{q\uparrow}0_{q\downarrow}...\rangle,\\
	(a~c_{q\uparrow}-b~c_{q\downarrow})|G\rangle &=& a|0_{q\uparrow}1_{q\downarrow}...\rangle+b|1_{q\uparrow}0_{q\downarrow}...\rangle,\\
	(a~c_{q\downarrow}-b~c_{q\uparrow})|G\rangle &=& -a|1_{q\uparrow}0_{q\downarrow}...\rangle-b|0_{q\uparrow}1_{q\downarrow}...\rangle,\\
	(a~c_{q\downarrow}+b~c_{q\uparrow})|G\rangle &=& -a|1_{q\uparrow}0_{q\downarrow}...\rangle+b|0_{q\uparrow}1_{q\downarrow}...\rangle.
	\end{eqnarray}
	The annihilation of an electron with momentum $q$ and spin $\sigma$ can be thought of as the creation of a hole with momentum $-q$ and spin $-\sigma$ w.r.t the quasiparticle vacuum $|0'\rangle$,
	\begin{equation}|0_{q\uparrow}1_{q\downarrow}...\rangle =h^{\dagger}_{-q,\downarrow}|0'\rangle,\hspace{2cm}|1_{q\uparrow}0_{q\downarrow}...\rangle = h^{\dagger}_{-q,\uparrow}|0'\rangle.\end{equation}
	We also defined,  \begin{eqnarray}\label{eq:triplet}\hat{\beta}_{\pm} &=& \frac{1}{2}(d^{\dagger}_{k\uparrow}d^{\dagger}_{q\uparrow}\pm d^{\dagger}_{k\downarrow}d^{\dagger}_{q\downarrow}),\\\label{eq:singlet}\hat{\beta'}_{\pm} &=& \frac{1}{2}(d^{\dagger}_{k\uparrow}d^{\dagger}_{q\downarrow}\pm d^{\dagger}_{k\downarrow}d^{\dagger}_{q\uparrow}),\end{eqnarray} to be the operators that correspond to entangled electron-electron pairs in the condensate, and \begin{equation}|\phi_{h}\rangle = a~h_{-q\uparrow}^{\dagger}|0'\rangle+b~h_{-q\downarrow}^{\dagger}|0'\rangle.\end{equation}
	
	Assuming that $\delta k$ is along $z$ direction, we note that the two electrons forming a spin singlet with opposite momenta in the transverse direction is described by the state $\hat{\beta'}_{-}|0\rangle$, with the center of mass $R$ moving along the direction of $\delta k$~\cite{spintron}:
	\begin{equation}
	\label{singlet}
	2\langle r_{1}, r_{2}|~\hat{\beta'}_{-}|0\rangle =e^{2i\delta k R}\sum\limits_{\kappa=\pm k_{F}}e^{i\kappa(r_{1}-r_{2})}\frac{1}{\sqrt{2}}(|\uparrow\downarrow\rangle-|\downarrow\uparrow\rangle)=\langle r_{1}, r_{2}|~e^{2i\delta k R}~	\sum\limits_{\pm\kappa}d_{\kappa\uparrow}^{\dagger}d_{-\kappa\downarrow}^{\dagger}|0\rangle = \langle r_{1}, r_{2}|~e^{2i\delta k R}~\hat{N}_{\kappa}^{\dagger}|0\rangle.\end{equation}The second equality follows from changing ($\kappa\rightarrow-\kappa$) in the sum term with the negative sign. We call the operator $\hat{N}_{\kappa}^{\dagger}$ a Cooper pair creation operator, introduced in Appendix~\ref{Ap:bcscoherent}. A superconductor imposing the final state boundary condition can hence, in principle, deterministically transfer the quantum information from the incoming electron to the outgoing hole. We stress here that there is no superluminal transfer of quantum information happening inside the superconductor, as the limiting factor for information transfer within the superconductor is the velocity of phonons that mediate the pairing interaction. The fact that all the interactions involved are local, and the pairing interaction within the superconductor only creates entangled singlet pairs, makes Andreev reflections a unitary process analogous to the final state projection models for black hole evaporation~\cite{projection,horowitz}. This is clearly different from methods of achieving deterministic transfer of quantum information using ordinary teleportation, by first doing a Bell measurement and then selecting out only the favorable measurement outcomes via a postselection~\cite{lloyd2011closed}. 
	
	\section{BCS ground state as a coherent state\label{Ap:bcscoherent}}
	Here, we review that the superconducting ground state can be seen as a coherent state of a Cooper pair creation operator that has a form identical to the one in Eq.~\eqref{singlet}. The phase of the coherent state is same as the phase of the BCS ground state. To see this, we rewrite the BCS ground state~\cite{annett2004superconductivity},
	\begin{eqnarray}
	|\Psi_{BCS}\rangle = 		\prod_{k}(u_{k}+v_{k}e^{-i\chi} d_{k\uparrow}^{\dagger}d_{-k\downarrow}^{\dagger})|0\rangle&=&\prod_{k}u_{k}(1+\frac{v_{k}}{u_{k}}e^{-i\chi} d_{k\uparrow}^{\dagger}d_{-k\downarrow}^{\dagger})|0\rangle\\
	&=& \prod_{k}u_{k}e^{\frac{v_{k}}{u_{k}}e^{-i\chi} d_{k\uparrow}^{\dagger}d_{-k\downarrow}^{\dagger}}|0\rangle\\
	&=& (\prod_{k}u_{k})e^{\sum\limits_{k}\frac{v_{k}}{u_{k}}e^{-i\chi} d_{k\uparrow}^{\dagger}d_{-k\downarrow}^{\dagger}}|0\rangle\\
	&=&U~e^{(\sum\limits_{k}\hat{N}_{k}^{\dagger})e^{-i\chi}}|0\rangle = U~e^{\hat{N}^{\dagger}e^{-i\chi}}|0\rangle.
	\end{eqnarray}
	Here $U$ is an overall normalization constant. The operator  $\hat{N_{k}}^{\dagger}$ is a Cooper pair creation operator:
	\begin{equation}
	\hat{N}^{\dagger}|0\rangle =\sum\limits_{k}\hat{N}_{k}^{\dagger}|0\rangle = \sum\limits_{k}\frac{v_{k}}{u_{k}} d_{k\uparrow}^{\dagger}d_{-k\downarrow}^{\dagger}|0\rangle =  \sum\limits_{k}g(k) d_{k\uparrow}^{\dagger}d_{-k\downarrow}^{\dagger}|0\rangle.	
	\end{equation}
	Observe that  $\hat{N}^{\dagger}$ creates a spin singlet state and $g(k)$ is the amplitude of the two particle wave function in the Fourier space. For \textit{s wave} superconductors $g(k)$ is symmetric, $g(k) = g(|k|)$. We included the sum over both $k$ and $-k$ pairs in the definition of the Cooper pair creation operator $\hat{N}_{k}$ -- labeled by $k$ -- to make the spherical symmetry of the condensate wavefunction explicit. Further, requiring that a new pair of quasiparticles entering the condensate also has this symmetry ensures that the symmetry remains unchanged as the condensate grows/evaporates.
	\bibliographystyle{ieeetr}
	\bibliography{ref}
\end{document}